\documentclass{eptcs-modified}

\usepackage{amsmath}
\usepackage{amssymb}
\usepackage{url}
\usepackage{amsthm}
\usepackage{graphicx}
\usepackage{amscd}

\usepackage{pgf} % added by SLR
\usepackage{tikz} % added by SLR

\begin{document}

\theoremstyle{definition}
\newtheorem{theorem}{Theorem}
\newtheorem*{theorem*}{Theorem}
\newtheorem{definition}[theorem]{Definition}
\newtheorem{problem}[theorem]{Problem}
\newtheorem{assumption}[theorem]{Assumption}
\newtheorem{corollary}[theorem]{Corollary}
\newtheorem{proposition}[theorem]{Proposition}
\newtheorem{example}[theorem]{Example}
\newtheorem{lemma}[theorem]{Lemma}
\newtheorem{observation}[theorem]{Observation}
\newtheorem{fact}[theorem]{Fact}
\newtheorem{question}[theorem]{Open Question}
\newtheorem{conjecture}[theorem]{Conjecture}
\newtheorem{addendum}[theorem]{Addendum}
\newcommand{\uint}{{[0, 1]}}
\newcommand{\Cantor}{{\{0,1\}^\mathbb{N}}}
\newcommand{\Baire}{{\mathbb{N}^\mathbb{N}}}
\newcommand{\name}[1]{\textsc{#1}}
\newcommand{\id}{\textrm{id}}
\newcommand{\dom}{\operatorname{dom}}
\newcommand{\win}{\textrm{Win}}
\newcommand{\NE}{\textrm{NE}}
\newcommand{\Det}{\textrm{Det}}
\newcommand{\C}{\textrm{C}}
\newcommand{\spe}{\textrm{SPE}}
\newcommand{\lpo}{\textrm{LPO}}
\newcommand{\llpo}{\textrm{LLPO}}
\newcommand{\lem}{\textrm{LEM}}
\newcommand{\mto}{\rightrightarrows}
\newcommand{\Sierp}{Sierpi\'nski }
\newcommand{\leqW}{\leq_{\textrm{W}}}
\newcommand{\leW}{<_{\textrm{W}}}
\newcommand{\equivW}{\equiv_{\textrm{W}}}
\newcommand{\geqW}{\geq_{\textrm{W}}}
\newcommand{\pipeW}{|_{\textrm{W}}}

\newcommand{\hide}[1]{}

\title{Weihrauch degrees of finding equilibria in sequential games\thanks{An extended abstract of this work has appeared in the Proceedings of CiE 2015 \cite{paulyleroux3-cie}.}}

\author{
St\'ephane Le Roux
\institute{D\'epartement d'informatique\\ Universit\'e libre de
Bruxelles, Belgique}
\email{Stephane.Le.Roux@ulb.ac.be}
\and
Arno Pauly
\institute{Computer Laboratory\\ University of Cambridge, United Kingdom}
\email{Arno.Pauly@cl.cam.ac.uk}
}

\def\titlerunning{Weihrauch degrees of finding equilibria}
\def\authorrunning{S. Le Roux \& A. Pauly}
\maketitle

\begin{abstract}
We consider the degrees of non-computability (Weihrauch degrees) of finding winning strategies (or more generally, Nash equilibria) in infinite sequential games with certain winning sets (or more generally, outcome sets). In particular, we show that as the complexity of the winning sets increases in the difference hierarchy, the complexity of constructing winning strategies increases in the effective Borel hierarchy.
\end{abstract}

\section{Overview}
We consider questions of (non)computability related to infinite sequential games played by any countable number of players. The best-known example of such games are Gale-Stewart games \cite{gale2}, which are two-player win/lose games. The existence of winning strategies in (special cases of) Gale-Stewart games is often employed to show that truth-values in certain logics are well-determined. The degrees of noncomputability of variations of (Borel) determinacy \cite{martin} can be studied using our techniques, and several are fully classified.

This work falls within the research programme to study the computational content of mathematical theorems in the Weihrauch lattice, which was outlined by \name{Brattka} and \name{Gherardi} in \cite{brattka3}. In particular, it continues the investigation of the Weihrauch degrees of operations mapping games to their equilibria started in \cite{paulyincomputabilitynashequilibria}. There, finding pure and mixed Nash equilibria in two-player games with finitely many actions in strategic form were classified.

One motivation for this line of inquiry is the general stance that solution concepts in game theory can only be convincing if the players are capable of (at least jointly) computing them, taken e.g.~in \cite{paulyphd}. Even if we allow for some degree of hypercomputation, or are, e.g.,~willing to tacitly replace actually attaining a solution concept by some process (slowly) converging to it, we still have to reject solution concepts with too high a Weihrauch degree.

The results for determinacy of specific pointclasses that we provide are a refinement of results obtained in reverse mathematics by \name{Nemoto}, \name{MedSalem} and \name{Tanaka} \cite{nemoto}; the first is also a uniformization of a result by \name{Cenzer} and \name{Remmel} \cite{cenzer}. For some represented pointclass $\Gamma$, let $\Det_\Gamma : \Gamma \mto \Cantor$ be the map taking a $\Gamma$-subset $A$ of Cantor space to a (suitably encoded) Nash equilibrium in the sequential two-player game with alternating moves where the first player wins if the induced play is in $A$, and the second player wins otherwise. Let $\mathcal{A}$ be the closed subsets of Cantor space, and $\mathfrak{D} := \{U \setminus U' \mid U, U' \in \mathcal{A}\}$. Some of our results are:
\begin{theorem*}
$\Det_{\mathcal{A}} \equivW \C_\Cantor$ and $\Det_{\mathfrak{D}} \equivW \C_\Cantor \star \lim$.
\end{theorem*}

We have two remarks. One, by combining the preceding theorem with the main result of \cite{gherardi4}, we find that $\Det_\mathfrak{D}$ is equivalent to the Bolzano-Weierstrass-Theorem. This may be a bit unexpected in particular seeing that $\C_\Cantor \star\lim$ is not (yet) known to contain a plethora of mathematical theorems (unlike, e.g.,~$\C_\Cantor$). Two, we already need to use a limit operator in order to move up one level of the difference hierarchy -- rather than being able to move up one level in the Borel hierarchy as one may have expected naively. Thus, this observation may complement Harvey Friedman's famous result \cite{friedman2} that proving Borel determinacy requires repeated use of the axiom of replacement.

Another group of results is based on inspecting the various results extending Borel determinacy to more general classes of games (and solution concepts) in \cite{leroux3,leroux4,paulyleroux2,paulyleroux2-arxiv}. If we instantiate these generic results with specific determinacy version as above, we can prove for some of them that they are actually optimal w.r.t.~Weihrauch reducibility. We shall state two such classifications explicitly.

Consider two-player sequential games with finitely many \emph{outcomes} and antagonistic (inverse of each other) linear preferences over the outcomes. For any upper set of outcomes w.r.t.~some player's preference let the corresponding set of plays be open or closed. Let $\NE^{ap}_{\mathcal{O}\cup\mathcal{A}}$ be the operation taking such a game (suitably encoded) and producing a Nash equilibrium. Then:
\begin{theorem*}
$\NE^{ap}_{\mathcal{O}\cup\mathcal{A}} \equivW \C_\Cantor \times \lpo^*$
\end{theorem*}

Those games will even have subgame-perfect equilibria, and we let $\spe_{\mathcal{O}\cup\mathcal{A}}$ be the operation mapping such games to a subgame-perfect equilibrium.

\begin{theorem*}
$\spe_{\mathcal{O}\cup\mathcal{A}} \equivW \lim$
\end{theorem*}

Various further classifications are obtained, and adhere to the scheme that algebraic combinations of very common Weihrauch degrees appear, which is already exhibited by our examples above.

\section{Fundamentals}
\subsection{Background on represented spaces}
We briefly recall some fundamental concepts on represented spaces following \cite{pauly-synthetic-arxiv}, to which the reader shall also be referred for a more detailed presentation. The concept behind represented spaces essentially goes back to \name{Weihrauch} and \name{Kreitz} \cite{kreitz}, the name may have first been used by \name{Brattka} \cite{brattka13}. A \emph{represented space} is a pair $\mathbf{X} = (X, \delta_X)$ of a set $X$ and a partial surjection $\delta_X : \subseteq \Baire \to X$. A function between represented spaces is a function between the underlying sets. For $f : \mathbf{X} \to \mathbf{Y}$ and $F : \subseteq \Baire \to \Baire$, we call $F$ a realizer of $f$ (notation $F \vdash f$), iff $\delta_Y(F(p)) = f(\delta_X(p))$ for all $p \in \dom(f\delta_X)$, i.e.~if the following diagram commutes:
 $$\begin{CD}
\Baire @>F>> \Baire\\
@VV\delta_\mathbf{X}V @VV\delta_\mathbf{Y}V\\
\mathbf{X} @>f>> \mathbf{Y}
\end{CD}$$
A map between represented spaces is called computable (continuous), iff it has a computable (continuous) realizer. Similarly, we call a point $x \in \mathbf{X}$ computable, iff there is some computable $p \in \Baire$ with $\delta_\mathbf{X}(p) = x$. A priori, the notion of a continuous map between represented spaces and a continuous map between topological spaces are distinct and should not be confused!

Given two represented spaces $\mathbf{X}$, $\mathbf{Y}$ we obtain a third represented space $\mathcal{C}(\mathbf{X}, \mathbf{Y})$ of continuous functions from $X$ to $Y$ by letting $0^n1p$ be a $[\delta_X \to \delta_Y]$-name for $f$, if the $n$-th Turing machine equipped with the oracle $p$ computes a realizer for $f$. As a consequence of the UTM theorem, $\mathcal{C}(-, -)$ is the exponential in the category of continuous maps between represented spaces, and the evaluation map is even computable (as are the other canonic maps, e.g.~currying).

This function space constructor, together with two represented spaces, $\mathbb{N} = (\mathbb{N}, \delta_\mathbb{N})$ and $\mathbb{S} = (\{\bot, \top\}, \delta_\mathbb{S})$, allows us to obtain a model of \name{Escard\'o}'s synthetic topology \cite{escardo}. The representation are given by $\delta_\mathbb{N}(0^n10^\mathbb{N}) = n$, $\delta_\mathbb{S}(0^\mathbb{N}) = \bot$ and $\delta_\mathbb{S}(p) = \top$ for $p \neq 0^\mathbb{N}$. It is straightforward to verify that the computability notion for the represented space $\mathbb{N}$ coincides with classical computability over the natural numbers. The \Sierp space $\mathbb{S}$ in turn allows us to formalize semi-decidability. The computable functions $f : \mathbb{N} \to \mathbb{S}$ are exactly those where $f^{-1}(\{\top\})$ is recursively enumerable (and thus $f^{-1}(\{\bot\})$ co-recursively enumerable).

In general, for any represented space $\mathbf{X}$ we obtain two spaces of subsets of $\mathbf{X}$; the space of open sets $\mathcal{O}(\mathbf{X})$ by identifying $f \in \mathcal{C}(\mathbf{X}, \mathbb{S})$ with $f^{-1}(\{\top\})$, and the space of closed sets $\mathcal{A}(\mathbf{X})$ by identifying $f \in \mathcal{C}(\mathbf{X}, \mathbb{S})$ with $f^{-1}(\{\bot\})$. The properties of the spaces of open and closed sets, namely computability of the usual operations, follow from a few particular computable functions on \Sierp space $\mathbb{S}$ and the fundamental function space properties.

We require further classes of sets (often called pointclasses in this context) as represented spaces. A general approach to this is found in synthetic descriptive set theory suggested in \cite{pauly-descriptive}. As shown in \cite{pauly-gregoriades-arxiv}, that approach yields consistent definitions with the ones used in \name{Moschovakis}' effective descriptive set theory \cite{moschovakis}. Here, we shall just directly provide representations that suffice. Some definitions essentially already appeared in \cite{brattka} and/or \cite{selivanov5}.

\begin{definition}
Given a represented pointclass $\Gamma$, we represent $\{A^C \mid A \in \Gamma\}$ by reinterpreting a name for $A$ as a name for $A^C$. Furthermore, we represent $\{\bigcup_{i \in \mathbb{N}} A_i \mid \forall i \in \mathbb{N} \ A_i \in \Gamma\}$ by identifying suitable sequences in $\Gamma^\mathbb{N}$. Given represented pointclasses $\Gamma_1, \Gamma_2$, we represent $\{A \setminus B \mid A \in \Gamma_1, B \in \Gamma_2\}$ by identifying the suitable pairs in $\Gamma_1 \times \Gamma_2$.
\end{definition}

Note that only the first of the three constructions preserves admissibility of the representations.

We will make use of the jump of a represented space. This is based on the limit operator $\lim : \subseteq \Baire \to \Baire$ defined via $\lim(p)(n) = \lim_{i \to \infty} p(\langle n, i\rangle)$. We then define $(X, \delta_X)' = (X, \delta_X \circ \lim)$. This is iterated along $\mathbf{X}^{(0)} := \mathbf{X}$ and $\mathbf{X}^{(n+1)} := (\mathbf{X}^{(n)})'$. In \cite{gherardi4}, the jump was extended to multivalued functions via $(f : \mathbf{X} \mto \mathbf{Y})^{[n]} := (f : \mathbf{X}^{[n]} \mto \mathbf{Y})$.

\subsection{Informal background on infinite sequential games}

We use the formal definitions of sequential games and related concepts from \cite{paulyleroux2,paulyleroux2-arxiv} and \cite{leroux3}. Informally, given a fixed (w.l.o.g.) set $C$, we let the players sequentially choose elements in $C$ until an infinite sequence in $C^{\omega}$ is generated. Whose turn it is depends on the finite history of choices. The outcome (from a set $O$) of the game depends on the generated sequence in $C^{\omega}$, and each player may compare outcomes via a binary relation over $O$, called preference. A strategy of a player is an object that fully specifies what the choice of the player would be for each possible finite history that requires this player to play. A combination of one strategy per player is called a strategy profile and it induces one unique infinite sequence in $C^{\omega}$, and thus one unique outcome. So, preferences may be lifted from outcomes to strategy profiles. A Nash equilibrium is a profile such that no player can unilaterally change strategies and induce a (new) outcome that he or she prefers over the old one. We also consider a refinement of the concept of a Nash equilibrium, namely subgame-perfect Nash equilibria. Intuitively, a strategy profile is subgame-perfect, if it still forms an equilibrium if the game were started at an arbitrary history.

\subsection{Background on infinite sequential games}
Our presentation of the required background on infinite sequential games is modelled closely on the corresponding section in \cite{paulyleroux2-arxiv}, and on \cite{leroux3}.

It is convenient to introduce the abstract notion of a \emph{game} first, before introducing additional structure later on. A game is a tuple $\langle A, (S_a)_{a \in A}, (\prec_a)_{a \in A}\rangle$ consisting of a non-empty set $A$ of \emph{agents} or \emph{players}, for each agent $a \in A$ a non-empty set $S_a$ of \emph{strategies}, and for each agent $a \in A$ a \emph{preference} relation $\mathalpha{\prec}_a \subseteq \left (\prod_{a \in A} S_a \right ) \times \left (\prod_{a \in A} S_a \right )$. The generic setting suffices to introduce the notion of a Nash equilibrium: A \emph{strategy profile} $\sigma \in \left (\prod_{a \in A} S_a \right )$ is called a Nash equilibrium, if for any agent $a \in A$ and any strategy $s_a \in S_a$ we find $\neg \left (\sigma \prec_a \sigma_{a\mapsto s_a} \right )$, where $\sigma_{a\mapsto s_a}$ is defined by $\sigma_{a\mapsto s_a}(b) = \sigma(b)$ for $b \in A \setminus \{a\}$ and $\sigma_{a\mapsto s_a}(a) = s_a$. In words, no agent prefers over a Nash equilibrium some other situation that only differs in her choice of strategy.

\begin{definition}[Infinite sequential game, {cf.~\cite[Definition 1.1]{leroux3}}]\label{defn:ifg} An
  \emph{infinite sequential game}  is an object $\langle A,C,d,O,v,(\prec_a)_{a\in A}\rangle$ complying with the following.
\begin{enumerate}
\item $A$ is a non-empty set (of agents).
\item $C$ is a non-empty set (of choices).
\item $d:C^*\to A$ (assigns a decision maker to each stage of the game).
\item $O$ is a non-empty set (of possible outcomes of the game).
\item $v:C^\omega\to O$ (uses outcomes to value the infinite sequences of choices).
\item Each $\prec_a$ is a binary relation over $O$ (modelling the preference of agent $a$).
\end{enumerate}
\end{definition}

The intuition behind the definition is that agents take turns to make a choice. Whose turn it is depends on the past choices via the function $d$ (which often will be just alternating play). Over time, the agents thus jointly generate some infinite sequence, which is mapped by $v$ to the outcome of the game. Note that using a single set of actions $C$ for each step just simplifies the notation; a generalization to varying action sets is straightforward. In the present paper, typically $A$ and $O$ will be countable, and $C$ even finite.

The infinite sequential games are linked to abstract games as follows: the agents remain the agents and the strategies of agent $a$ are the functions $s_a : d^{-1}(\{a\}) \to C$. We can then safely regard a strategy profile as a function $\sigma : C^* \to C$ whose \emph{induced play} is defined below, where for an infinite sequence $p \in C^\omega$ we let $p_n$ be its $n$-th value, and $p_{\leq n} = p_{<n + 1} \in C^*$ be its finite prefix of length $n$.
 \begin{definition}[Induced play and outcome, {cf.~\cite[Definition 1.3]{leroux3}}]\label{defn:ipo}
Let $s:C^*\to C$ be a strategy profile. The \emph{play $p=p^{\lambda}(s)\in
  C^\omega$  induced by $s$ starting at $\lambda \in C^*$} is defined inductively through its prefixes:
$p_n = \lambda_n$ for $n \leq |\lambda|$ and $p_{n}:=s(p_{< n})$ for $n > |\lambda|$. Also, $v\circ p_{\lambda}(s)$ is the \emph{outcome induced by $s$ starting at $\lambda$}. The play (resp. outcome) induced by $s$ is the play (resp. outcome) induced by $s$ starting at $\varepsilon$.
\end{definition}

In the usual way to regard an infinite sequential game as a special abstract game, an agent prefers a strategy profile $\sigma$ to $\sigma'$, iff he prefers the outcome induced by $\sigma$ to the outcome induced by $\sigma'$. And indeed we shall call a strategy profile of an infinite sequential game a Nash equilibrium, iff it is a Nash equilibrium with these preferences. In a certain notation overload, we will in particular use the same symbols for the preferences over strategy profiles and the preferences over outcomes.

We also consider a refinement of the concept of a Nash equilibrium, namely subgame-perfect Nash equilibria. Intuitively, a strategy profile is subgame-perfect, if it still forms an equilibrium if the game were started in an arbitrary position.

As a important special case we consider win/lose games. These are games with two players $a, b$ and two outcomes $w_a$, $w_b$, where $a$ prefers $w_a$ to $w_b$ and $b$ prefers $w_b$ to $w_a$. We say that $a$ wins the game, if outcome $w_a$ is reached, and call the set of all plays that induce outcome $w_a$ as the winning set for $a$ (likewise for $b$ and $w_b$).

\subsection{Background on Weihrauch reducibility}
A few years ago several authors (\name{Gherardi} and \name{Marcone} \cite{gherardi}, \name{P.} \cite{paulyreducibilitylattice,paulyincomputabilitynashequilibria}, \name{Brattka} and \name{Gherardi} \cite{brattka3}) noticed that a reducibility notion based on previous work by \name{Weihrauch} \cite{weihrauchb,weihrauchc} would provide a very interesting setting for a metamathematical inquiry into the computational content of mathematical theorems. The fundamental research programme was outlined in \cite{brattka3}, and the introduction in \cite{hoelzl} may serve as a recent survey.

\label{subsec:weihrauch}
\begin{definition}[Weihrauch reducibility]
\label{def:weihrauch}
Let $f,g$ be multi-valued functions on represented spaces.
Then $f$ is said to be {\em Weihrauch reducible} to $g$, in symbols $f\leqW g$, if there are computable
functions $K,H:\subseteq\Baire\to\Baire$ such that $K\langle \id, GH \rangle \vdash f$ for all $G \vdash g$.
\end{definition}
The relation $\leqW$ is reflexive and transitive. We use $\equivW$ to denote equivalence regarding $\leqW$,
and by $\leW$ we denote strict reducibility. By $\mathfrak{W}$ we refer to the partially ordered set of equivalence classes. As shown in \cite{paulyreducibilitylattice,brattka2}, $\mathfrak{W}$ is a distributive lattice, and also the usual product operation on multivalued function induces an operation $\times$ on $\mathfrak{W}$. The algebraic structure on $\mathfrak{W}$ has been investigated in further detail in \cite{paulykojiro,paulybrattka4}.

There are two relevant unary operations defined on $\mathfrak{W}$, both happen to be closure operators. The operation $^*$ was introduced in \cite{paulyreducibilitylattice,paulyincomputabilitynashequilibria} by setting $f^0 := \id_\Baire$, $f^{n+1} := f \times f^{n}$ and then $f^*(n,x) := f^n(x)$. It corresponds to making any finite number of parallel uses of $f$ available. Similarly, the \emph{parallelization} operation $\widehat{\phantom{f}}$ from \cite{brattka2,brattka3} makes countably many parallel uses available by $\widehat{f}(x_0, x_1, x_2, \ldots) := (f(x_0), f(x_1), f(x_2), \ldots)$.

We will make use of an operation $\star$ defined on $\mathfrak{W}$ that captures aspects of function composition. Following \cite{gherardi4,paulybrattka3cie}, let $f \star g := \max_{\leqW} \{f_0 \circ g_0 \mid f \equivW f_0 \wedge g \equivW g_0\}$. We understand that the quantification is running over all suitable functions $f_0$, $g_0$ with matching types for the function composition. It is not obvious that this maximum always exists, this is shown in \cite{paulybrattka4} using an explicit construction for $f \star g$. Like function composition, $\star$ is associative but generally not commutative. We use $\star$ to introduce iterated composition via setting $f^{(0)} := \id_\Baire$ and $f^{(n+1)} = f^{(n)} \star f$.

All computable multivalued functions with a computable point in their domain are Weihrauch equivalent, this degree is denoted by $1$.

An important source for examples of Weihrauch degrees that are relevant in order to classify theorems are the closed choice principles studied in e.g.~\cite{brattka3,paulybrattka}:
\begin{definition}
Given a represented space $\mathbf{X}$, the associated closed choice principle $\C_\mathbf{X}$ is the partial multivalued function $\C_\mathbf{X} : \subseteq \mathcal{A}(\mathbf{X}) \mto \mathbf{X}$ mapping a non-empty closed set to an arbitrary point in it.
\end{definition}

For any uncountable compact metric space $\mathbf{X}$ we find that $\C_\mathbf{X} \equivW \C_\uint$. For well-behaved spaces, using closed choice iteratively does not increase its power, in particular $\C_\mathbb{N} \star \C_\mathbb{N} \equivW \C_\mathbb{N}$ and $\C_\uint \star \C_\uint \equivW \C_\uint$. Likewise, it was shown that $\C_{\mathbb{R}^n} \equivW \C_{\mathbb{R}^n} \star \C_{\mathbb{R}^n} \equivW \C_\mathbb{N} \times \C_\uint \equivW \C_\mathbb{N} \star \C_\uint \equivW \C_\uint \star \C_\mathbb{N}$ for any $n > 0$. Closed choice for $\uint$ and $\Cantor$ is incomparable. Furthermore, $\C_\Cantor \equivW \widehat{\C_\Cantor} \equivW \widehat{\C_{\{0,1\}}}$. The degree $\C_\uint$ is closely linked to $\textrm{WKL}$ in reverse mathematics, while $\C_\mathbb{N}$ is Weihrauch-complete for functions computable with finitely many mindchanges.

Another typical degree is $\lpo$, which has important representatives such as $\mathalpha{\neg} : \mathbb{S} \to \mathbb{S}$, the characteristic function of $0^\mathbb{N}$, the characteristic function of $0$ in $\mathbb{R}$, $\mathalpha{\neq} : \Cantor \times \Cantor \to \{0,1\}$ and $\operatorname{IsEmpty} : \mathcal{A}(\Cantor) \to \{0,1\}$.

 Furthermore, we require the degree obtained from the limit operator $\lim : \subseteq \Baire \to \Baire$. This degree was studied  by \name{von Stein} \cite{stein}, \name{Mylatz} \cite{Mylatz} and \name{Brattka} \cite{brattka11,brattka}, with the latter noting in \cite{brattka} that it is closely connected to the Borel hierarchy. \name{Hoyrup}, \name{Rojas} and \name{Weihrauch} have shown that $\lim$ is equivalent the Radon-Nikodym derivative in \cite{hoyrup2b}; while \cite{pauly-fouche} by \name{P.} and \name{Fouch\'e} exhibited some constructions of Radon measures that are equivalent to $\lim$. The degree $\lim$ also appears in the context of models of hypercomputation as shown by \name{Ziegler} \cite{ziegler2,ziegler7}, and captures precisely the additional computational power that certain solutions to general relativity could provide beyond computability \cite{hogarth}. It is related to the examples above via $\C_\mathbb{N} \times \C_\uint \leW \lim \equivW \lim \times \lim \leW \lim \star \lim$, and $\lpo \leW \lpo^* \leW \C_\mathbb{N} \leW \lim \equivW \widehat{\lpo}$. The (strict) hierarchy $(\lim^{(n)})_{n \in \mathbb{N}}$ plays a very similar role in the Weihrauch degrees as the iterated Halting problems fill in the Turing degrees. We always find $f^{[n]} \leqW f \star \lim^{(n)}$, and for many natural functions $f$, equivalence holds, since $\left ( f \times \id \right) ^{[n]} \equivW f \star \lim^{(n)}$.

\subsection{Defining the problems of interest}
Let $\Gamma$ be a represented pointclass over $\Cantor$. In a straightforward fashion, we can obtain a representation of the infinite sequential games with countably many agents, countably many outcomes, sets of choices $C = \{0,1\}$ and $\Gamma$-measurable valuation function $v : \Cantor \to O$. The representation encodes the number of agents and outcomes available, for each upper set of outcome the $\Gamma$-set of plays resulting in it, the map $d$ as a look-up table, and the relations $\prec_a$ as look-up tables. We always assume that the inverse of any preference relation is well-founded (this guarantees that equilibria exist). Using a canonic isomorphism $\{0,1\}^* \cong \mathbb{N}$, we will pretend that the space of strategy profiles in such a game is $\Cantor$.

We now consider the following multivalued functions:
\begin{enumerate}
\item $\Det_\Gamma$ takes a two-player win/lose game as input, where the first player has a winning set in $\Gamma$. Valid outputs are the Nash equilibria, i.e.~the pairs of strategies where one strategy is a winning strategy.
\item $\win_\Gamma$ has the same inputs as $\Det_\Gamma$, and decides which player (if any) has a winning strategy.
\item $\textrm{FindWS}_\Gamma$ is the restriction of $\Det_\Gamma$ to games where the first player has a winning strategy.
\item $\NE_\Gamma$ takes as input a game with countably many players, finitely many outcomes, and linear preferences, where each upper set of outcomes (w.r.t. each player preference) comes from a $\Gamma$-set. The valid outputs are the Nash equilibria.
\item $\NE_\Gamma^{\omega o}$ takes as input a game with countably many players, countably many outcomes, and linear preferences, where each upper set of outcomes (w.r.t. each player preference) comes from a $\Gamma$-set. The valid outputs are the Nash equilibria.
\item $\NE_\Gamma^{ap}$ is the restriction of $\NE_\Gamma$ to the two-player games with antagonistic preferences (i.e.~$\prec_a = \prec_b^{-1}$).
\item $\spe_\Gamma$ takes as input a two-player game with finitely many outcomes and antagonistic preferences, where each upper set of outcomes comes from a $\Gamma$-set. Valid outputs are the subgame perfect equilibria.
\end{enumerate}

We abbreviate $\overline{\Gamma} := \{U^C \mid U \in \Gamma\}$. Some trivial reducibilities between these problems are: $\win_{\Gamma} \equivW \win_{\overline{\Gamma}}$, $\Det_{\Gamma} \equivW \Det_{\overline{\Gamma}}$, $\textrm{FindWS}_\Gamma \leqW \Det_\Gamma \leqW \NE_{\Gamma \cup \overline{\Gamma}}^{ap} \leqW \spe_{\Gamma \cup  \overline{\Gamma}}$ and $\NE_\Gamma^{ap} \leqW \NE_\Gamma$.

Throughout the paper we assume that $\Gamma$ is determined (which implies that all operations are well-defined in the first place), closed under rescaling and finite intersection with clopens, and that $\emptyset, \Cantor \in \Gamma$. All such closure properties (including those appearing as conditions in the results) are assumed to hold in a uniformly computable way, e.g.~given a name for a set in $\Gamma$ and a clopen, we can compute a name for the intersection of the set with the clopen. With rescaling we refer to the operation $(w, A) \mapsto \{wp \mid p \in A\} : \{0,1\}^* \times \Gamma \to \Gamma$ and its inverse.

\subsection{The difference hierarchy}
The pointclasses we shall study in particular are the levels of the Hausdorff difference hierarchy. Intuitively, these are the sets that can be obtained as boolean combinations of open sets; and their level denotes the least complexity of a suitable term. Roughly following \cite[Section 22.E]{kechris}, we shall recall the definition of the difference hierarchy. We define a function $\textrm{par}$ from the countable ordinals to $\{0,1\}$ by $\textrm{par}(\alpha) = 0$, if there is a limit ordinal $\beta$ and a number $n \in \mathbb{N}$ such that $\alpha = \beta + 2n$; and $\textrm{par}(\alpha) = 1$ otherwise. For a fixed ordinal $\alpha$, we let $\mathfrak{D}_\alpha$ be the collection of sets $D$ definable in terms of a family $(U_\lambda)_{\lambda < \alpha}$ of open sets via: \[x \in D \Leftrightarrow \textrm{par}\left (\inf \{\beta \mid x \in U_\beta\} \right ) \neq \textrm{par}(\alpha)\]

In the preceding formula, we understand that $\inf \emptyset = \alpha$. In particular, $\mathfrak{D}_0 = \{\emptyset\}$ and $\mathfrak{D}_1 = \mathcal{O}$.

For our constructions, a different characterization is more useful, though: For some pointclass $\Gamma$, let $\mathfrak{D}(\Gamma) := \{\bigcup_{i \in I} v_iU_i \mid \forall i,j \in I v_i \in \{0,1\}^* \wedge U_i \in \Gamma \wedge v_i \nprec v_j\}$.

\begin{lemma}
\label{lemma:differencehierarchycharacterization}
$\mathfrak{D}_{\alpha+1} = \mathfrak{D}(\overline{\mathfrak{D}_\alpha})$ and, more generally, $\mathfrak{D}_{\alpha} = \mathfrak{D}\left (\overline{\bigcup_{\lambda < \alpha} \mathfrak{D}_\lambda} \right )$
\begin{proof}
The proof proceeds via induction. The base case is $\mathfrak{D}_1 = \mathcal{O} = \mathfrak{D}(\{\Cantor\})$, and straightforward. Now assume that the claim holds for all $\beta < \alpha$.

If $A \in \mathfrak{D}_{\alpha+1}$, then there is a witnessing family $(U_\beta)_{\beta < \alpha+1}$ of open sets. Let $U_\alpha = \bigcup_{i \in I} v_i\Cantor$ with a prefix free family $(v_i)_{i \in I}$. For $\lambda < \alpha$ and $i \in I$, we define an open set $U_\lambda^i := \{y \in \Cantor \mid v_iy \in U_\lambda\}$. Then let $A^i \in \mathfrak{D}_{\alpha}$ be the set constructed from the family $(U^i_\lambda)_{\lambda < \alpha}$, and let $U^i := \left ( A^i\right )^C$. Now $A = \bigcup_{i \in I} v_iU^i$ witnesses that $A \in \mathfrak{D}(\overline{\mathfrak{D}_\alpha})$.

If $A \in \mathfrak{D}(\overline{\mathfrak{D}_\alpha})$, then there are witnesses $(v_i)_{i \in I}$ and $(U^i)_{i \in I}$ with $U^i \in \overline{\mathfrak{D}_\alpha}$. The latter is in turn witnessed by families of open sets $(U^i_\beta)_{\beta < \alpha}$. For $\beta < \alpha$, let $U_\beta = \bigcup_{i \in I} v_iU^i_\beta$, this is an open set again. Additionally, let $U_\alpha = \bigcup_{i \in I} v_i\Cantor$. Now the family $(U_\lambda)_{\lambda < \alpha+1}$ witnesses that $A \in \mathfrak{D}_{\alpha+1}$.
\end{proof}
\end{lemma}

\begin{observation}
\label{obs:diffclosure}
If $A_n$ is in $\mathfrak{D}_{\alpha}$ for all $n\in \mathbb{N}$, so is $A := \cup_{n\in\mathbb{N}}0^n1A_n$.
\begin{proof}
For all $n$ the set $A_n$ can be written $\cup_{i\in I_n}v_{ni}A_{ni}$, where $v_{ni}\not\prec v_{nj}$ and $A_{ni}\in \mathfrak{D}_{\lambda_{ni}}$ for some $\lambda_{ni} < \alpha$, so $A = \cup_{n\in\mathbb{N}, i\in I_n}0^n1v_{ni}A_{ni}$, where $0^n1v_{ni} \not\prec 0^m1v_{mj}$.
\end{proof}
\end{observation}

\begin{corollary}
\label{corr:diffhierpluspoint}
If $B_n$ is in $\overline{\mathfrak{D}_{\alpha}}$ for all $n\in \mathbb{N}$, so is $B := \{0^{\mathbb{N}}\}\cup\bigcup_{n\in\mathbb{N}}0^n1B_n$.
\end{corollary}

A fundamental result on the difference hierarchy is the Hausdorff-Kuratowski theorem stating that $\bigcup_{\alpha < \omega_1} \mathfrak{D}_\alpha = \Delta^0_2$ (where $\omega_1$ is the smallest uncountable ordinal), see e.g.~\cite[Theorem 22.27]{kechris}.

\section{The computational content of some determinacy principles}
\label{sec:det}
We begin by classifying the simplest non-computable games, namely games where the first player wants to reach some closed set. This classification essentially is a uniform version of a result by \name{Cenzer} and \name{Remmel} \cite{cenzer}.
\begin{theorem}
\label{theo:openclosedne}
$\textrm{FindWS}_\mathcal{A} \equivW \Det_\mathcal{A} \equivW \C_\Cantor$.
\begin{proof}
\begin{description}
\item[$\C_\Cantor \leqW \textrm{FindWS}_\mathcal{A}$]
Given a closed subset $A \in \mathcal{A}(\Cantor)$, we can easily obtain the game where only player 1 moves, and player 1 wins iff the induced play falls in $A$. If $A$ is non-empty, then player 1 has a winning strategy: Play any infinite sequence in $A$.
\item[$\textrm{FindWS}_\mathcal{A} \leqW \Det_\mathcal{A}$] Trivial.
\item[$\Det_\mathcal{A} \leqW \C_\Cantor$] Given the open winning set of player 2, we can modify the game tree by ending the game once we know for sure that player 2 will win. Now the set of strategy profiles where either player 1 wins and player 2 cannot win, or player 2 wins and player 1 cannot prolong the game, is a closed set effectively obtainable from the game. Moreover, it is non-empty, and any such strategy profile is a Nash equilibrium.
\end{description}
\end{proof}
\end{theorem}

\begin{proposition}
\label{prop:openclosedwin}
$\win_\mathcal{A} \equivW \lpo$
\begin{proof}
This follows by combining the constructions from the preceding theorem with the fact that $\operatorname{IsEmpty} : \mathcal{A}(\Cantor) \to \{0,1\}$ is equivalent to $\lpo$.
\end{proof}
\end{proposition}

We can use the results for $\mathcal{A}$ as the base case for classifying the strength of determinacy for the difference hierarchy.

\begin{lemma}[\footnote{This is a generalization of the proof idea for \cite[Theorem 3.7]{nemoto} by \name{Nemoto}, \name{MedSalem} and \name{Tanaka}. \cite[Theorem 3.7]{nemoto} states that $\textrm{ACA}_0$ proves determinacy for $\mathfrak{D}(\Sigma^0_1)$.}]
\label{theo:differencestep}
$\Det_{\mathfrak{D}(\Gamma)} \leqW \C_\Cantor \star \widehat{\left (\Det_{\Gamma} \times \win_{\Gamma} \right )}$ and $\win_{\mathfrak{D}(\Gamma)} \leqW \lpo \star_s \widehat{\win_{\Gamma}}$.\footnote{Here, $\star_s$ denotes the analogue of $\star$ for \emph{strong Weihrauch reducibility} (cf~\cite{gherardi4}). Its precise definition is not important in the following, and the statement remains true if $\star_s$ is replaced by $\star$.}
\begin{proof}
Let the winning set of the first player in the original game be $\bigcup_{i \in \mathbb{N}} v_iU_i$. We use $\widehat{\Det_{\Gamma} \times \win_{\Gamma}}$ to find out who wins each of the games with winning sets $U_i$, and a Nash equilibrium for each such game. Let $v_{n_i}$ be the subsequence of the $v_n$ where the first player wins.

Now consider the game where the first player's winning set is the open set $U' = \bigcup_{i \in \mathbb{N}}v_{n_i}\Cantor$. The first player is winning this derived game, iff he is winning the original game. Thus, by Proposition \ref{prop:openclosedwin} the claim $\win_{\mathfrak{D}(\Gamma)} \leqW \lpo \star_s \widehat{\win_{\Gamma}}$ follows. Due to Theorem \ref{theo:openclosedne}, we can use $\C_{\Cantor}$ to find a Nash equilibrium of this game. Then we combine this Nash equilibrium with those of the subgames to get a Nash equilibrium of the entire game, by letting both players play their equilibrium strategy in any of the subgames rooted at a $v_n$, and above those, letting the first player try to reach some $v_{n_i}$ and the second to avoid them. This is the claim $\Det_{\mathfrak{D}(\Gamma)} \leqW \C_\Cantor \star \widehat{\Det_{\Gamma} \times \win_{\Gamma}}$.
\end{proof}
\end{lemma}

\begin{observation}
$\Det_{\bigcup_{n \in \mathbb{N}} \Gamma_n} \equivW \coprod_{n \in \mathbb{N}} \Det_{\Gamma_n}$ and $\win_{\bigcup_{n \in \mathbb{N}} \Gamma_n} \equivW \coprod_{n \in \mathbb{N}} \win_{\Gamma_n}$
\end{observation}

We will relate deciding the winner and finding a winning strategy for games induced by sets from some level of the difference hierarchy to the \emph{lessor limited principle of omniscience} and the \emph{law of excluded middle} for $\Sigma^0_n$-formulae of the corresponding level. These principles were studied in \cite{kohlenbach,gherardi4,kihara3} (among others). Let $\left (\Sigma_n^0-\llpo \right ): \subseteq \Cantor \times \Cantor \mto \{0,1\}$ be defined via $i \in \left (\Sigma_n^0-\llpo\right )(p_0, p_1)$ iff $\forall k_1 \exists k_2 \ldots \natural k_n \ p_i(\langle k_1, \ldots, k_n\rangle) = 1$ (where $\natural = \forall$ if $n$ is odd and $\natural = \exists$ otherwise). Let $\left (\Sigma_n^0-\lem\right ) : \Cantor \to \{0,1\}$ be defined via $\left (\Sigma_n^0-\lem\right )(p) = 1$ iff $\forall k_1 \exists k_2 \ldots \natural k_n \ p(\langle k_1, \ldots, k_n\rangle) = 1$ and $\left (\Sigma_n^0-\lem\right )(p) = 0$ otherwise. Then:
\begin{proposition}
\label{prop:sigmallpo}
$\left (\Sigma_{n+1}^0-\llpo\right ) \equivW \llpo^{[n]}$ and $\left (\Sigma_{n+1}^0-\lem\right ) \equivW \lpo^{[n]}$
\end{proposition}

\begin{lemma}
\label{lemma:diffhierconstruction}
$\widehat{\left (\Sigma_{n}^0-\llpo\right )} \leqW \Det_{\mathfrak{D}_n}$ and $\left (\Sigma_{n}^0-\lem\right ) \leqW \win_{\mathfrak{D}_n}$.
\begin{proof}
We will first describe the construction for $\left (\Sigma_{n}^0-\lem\right ) \leqW \win_{\mathfrak{D}_n}$, which will then be reused for the remaining claim. The game structure will only depend on the parameter $n$, but not on the actual input for $\left (\Sigma_{n}^0-\lem\right )$. This input acts only on the winning set.

The game works as follows: The second player may pick some $k_1 \in \mathbb{N}$, or refuse to play. If the second player picks a number, then the first player may pick $k_2 \in \mathbb{N}$ or refuse to play. This alternating choice continues until $k_{n-1}$ has been chosen, or a player refuses to pick. A player refusing to pick a number loses. If all numbers are picked, the winner depends on the input $p$ to $\Sigma_{n}^0-\lem$ as follows: If $n$ is even and $\exists k_n \ p(\langle k_1,\ldots,k_n\rangle) = 1$, then player 1 wins. If $n$ is odd, and $\exists k_n \ p(\langle k_1,\ldots,k_n\rangle) = 0$, then player 2 wins. Note that this always describes an open component $U_{\textnormal{picked}}$ of the winning set of the respective player.

Furthermore, note that the set of plays $U_j$ where a value for $k_j$ was chosen is always an open set. Now the condition that the second player refused to pick first is $U_1^C \cup (U_2 \cap U_3) \cup (U_4 \cap U_5) \cup \ldots$. This makes for a winning set in $\mathfrak{D}_n$, as required. If player $1$ has a winning strategy in the game, the answer to $\left (\Sigma_{n}^0-\lem\right )(p)$ is $1$, if player $2$ wins, it is $0$.

The game for the reduction $\widehat{\left (\Sigma_{n}^0-\llpo\right )} \leqW \Det_{\mathfrak{D}_n}$ adds two layers above the game discussed before. First, player $2$ picks an index $j \in\mathbb{N}$ of one of the input pairs $\langle \langle p_1^0, p_1^1\rangle, \langle p_2^0, p_2^1\rangle, \ldots, \rangle$ of $\widehat{\left (\Sigma_{n}^0-\llpo\right )}$, and loses the game if he refuses to pick. Then player $1$ picks $i \in \{0,1\}$, and they play the game above on $p_j^i$. The extra layers do not impact the complexity of the winning set, in particular since the first two natural numbers are both to be chosen by player $2$. The map $j \mapsto i$ extractable from player $1$'s winning strategy is a valid solution to $\widehat{\left (\Sigma_{n}^0-\llpo\right)}$.
\end{proof}
\end{lemma}

\begin{theorem}
\label{theo:differencehierarchy}
$\Det_{\mathfrak{D}_{n+1}} \equivW \C_{\Cantor}^{[n]}$ and $\win_{\mathfrak{D}_{n+1}} \equivW \lpo^{[n]}$.
\begin{proof}
Note that $\widehat{\llpo^{[n]}} \equivW \widehat{\llpo}^{[n]}$ and $\widehat{\lpo^{[n]}} \equivW \lim^{(n+1)}$. One direction of the equivalences is provided by Lemma \ref{lemma:diffhierconstruction} (while taking into consideration Proposition \ref{prop:sigmallpo}). The other direction is shown by induction. The base case is provided by Theorem \ref{theo:openclosedne} and Proposition \ref{prop:openclosedwin}. The induction step uses Lemma \ref{theo:differencestep}.
\end{proof}
\end{theorem}

The main obstacle for extending Theorem \ref{theo:differencehierarchy} to the transfinite levels of the difference hierarchy lies in the absence of a standard representation for the space of countable ordinals, which would be required to formulate the statement in the first place. This problem is (partially) addressed in \cite{pauly-ordinals}. In the mean time, we can give part of the result for the $\omega$-th level:

\begin{corollary}
$\Det_{\mathfrak{D}_\omega} \leqW \C_\Cantor \star \widehat{\left (\coprod_{n \in \mathbb{N}} \lim^{(n)}\right )}$ and $\win_{\mathfrak{D}_\omega} \leqW \lpo \star \widehat{\left (\coprod_{n \in \mathbb{N}} \lim^{(n)}\right )}$
\end{corollary}

Knowing the Weihrauch degree of a mapping entails some information about the Turing degrees of outputs relative to the Turing degrees of inputs, this was explored in e.g.~\cite{brattka2,gherardi4,hoelzl,paulyphd}. Thus, we can obtain the following corollaries:

\begin{corollary}
\label{corr:differencehierarchy1}
Any computable game with a winning condition in $\mathfrak{D}_{n+1}$ has a winning strategy $s$ such that $s'$ is computable relative to $\emptyset^{(n+1)}$, and there is a computable game of this type such that any winning strategy computes a PA-degree relative to $\emptyset^{(n)}$.
\end{corollary}

\begin{corollary}
\label{corr:differencehierarchy2}
Let $(G_i)_{i \in \mathbb{N}}$ be an effective enumeration of computable games with winning conditions in $\mathfrak{D}_{n+1}$, and define $w \in \Cantor$ via $w(i) = 1$ iff the first player has a winning strategy in $G_i$. Then $w \leq_{\textrm{T}} \emptyset^{(n+1)}$, and there is an enumeration $(G_i)_{i \in \mathbb{N}}$ such that $w \equiv_{\textrm{T}} \emptyset^{(n+1)}$.
\end{corollary}

\begin{corollary}
\label{corr:differencehierarchy3}
There is a $\Sigma^0_{n+1}$-measurable function mapping games with winning conditions in $\mathfrak{D}_n$ to winning strategies, but no $\Sigma^0_{n}$-measurable such function.
\begin{proof}
As shown in \cite{brattka}, the map $\lim^{(n)}$ is Weihrauch-complete for the (effectively) $\Sigma_{n+1}^0$-measurable functions. From Theorem \ref{theo:differencehierarchy} we may conclude that $\lim^{(n)} \leW \Det_{\mathfrak{D}_{n+1}} \leW \lim^{(n+1)}$.
\end{proof}
\end{corollary}
Before ending this section, we shall make explicit a feature of the constructions used above:

\begin{proposition}\label{prop:find-hat}
 Consider a pointclass $\Gamma$ that is closed under the operation $(A_n)_{n \in \mathbb{N}} \mapsto \left (\{01^\mathbb{N}\} \cup \bigcup_{n \in \mathbb{N}} 1^n0A_n \right )$. Then: \[\textrm{FindWS}_\Gamma \equivW \widehat{\textrm{FindWS}_\Gamma}\]
\begin{proof}
Given a sequence of games, we construct a single combined game as follows: The second player picks a natural number, and loses the game if he fails to do so. If a number is chosen, the players proceed to the game with the corresponding index. If the first player has winning strategies in all input games, he can win the combined game by playing their combination -- and only by doing so.
\end{proof}
\end{proposition}

\section{The complexity of equilibrium transfer}
\label{sec:transfer}
In \cite{leroux3,leroux4,paulyleroux2,paulyleroux2-arxiv}, various results were provided that transfer Borel determinacy (or, somewhat more general, determinacy for some pointclass), to prove the existence of Nash equilibria (and sometimes even subgame-perfect equilibria) in multi-player multi-outcome infinite sequential games. In this section, we shall inspect those constructions and extract Weihrauch reductions from them.

In \cite{leroux4}, the first author gave a very general construction that allows to extend determinacy of win/lose games to the existence of Nash equilibria for two-player games of the same type. For brevity, we only consider the strength of the toy example from \cite{leroux4} here:
\begin{theorem}[Equilibrium transfer]
\label{theo:equitransfer}
$\NE^{ap}_\Gamma \leqW \Det_\Gamma^* \times \win_\Gamma^*$.
\begin{proof}
For any upper set of outcomes (for either players preferences), we construct the win/lose derived game where that player wins, iff he enforces the set, and loses otherwise. There are finitely many such games, so we can use $\win_\Gamma^*$ to decide which are won and which are lost. As shown in \cite{leroux4}, there will be a combination of upper sets of outcomes for each player, such that if both players enforce their upper set, this forms a Nash equilibrium. We use $\Det_\Gamma^*$ to compute Nash equilibria for all derived games in parallel, and then simply select the suitable strategies.
\end{proof}
\end{theorem}

Techniques suitable for multiplayer sequential games were then introduced in \cite{leroux3}, again by the first author. The computational content differs somewhat depending on whether there are finitely many different outcomes, or countably many.
\begin{theorem}[Constructing Nash equilibria]
\label{theo:construct-NE}
$\NE_\Gamma \leqW \widehat{\win_{\Gamma}} \times
\widehat{\Det_{\Gamma}}$, \\ $\NE^{\omega o}_\Gamma \leqW (\lim \star
\widehat{\win_{\Gamma}}) \times \widehat{\Det_{\Gamma}}$
\begin{proof}
Let us first prove the first statement. Let us consider a game with finitely many players. (Considering countably many players would be possible, but it
would reduce to the finite case since there are only finitely many
possible preference relations.) For each node of the game and each upper interval of the
preference of the player owning the node (that is, for countably many
cases) let us do two things in parallel: on the one hand, invoke $\win_\Gamma$ and
ask who is winning the win/lose subgame rooted at the node where all
the opponents of the node owner team up and try to yield an outcome
outside of the given preference-upper interval; on the other hand, invoke
$\Det_\Gamma$ and obtain a Nash equilibrium for that same game. We claim
that this information suffices to computably build a Nash equilibrium for
the original game along the proof from~\cite{leroux3}:
indeed, the \emph{best guarantee}~\cite[Definition 2.5]{leroux3} of a
player, which is the smallest preference-upper interval that the player
can enforce, can be computed since we already know who wins the relevant
derived games thanks to $\win_\Gamma$; the existential witness
from~\cite[Lemma 2.6]{leroux3} is a Nash equilibrium that has been already
computed by $\Det_\Gamma$; which is enough to deepen the guarantee as
in~\cite[Definition 2.7]{leroux3} and build the strategy profile
$\sigma$~\cite[Lemma 2.8]{leroux3}; and finally, the threats
that~\cite[Theorem 2.9]{leroux3} attaches along the play of $\sigma$ are
given by the Nash equilibria that have been already computed by $\Det_\Gamma$.

Let us now consider the similar, second statement. At each node, the associate games are now countably many, so knowing who wins each of them no longer suffices to compute the best guarantee of the node owner. To this purpose we use one $\lim$ operator per node, and the rest follows as in the finite case above, so $\NE^{\omega o}_\Gamma \leqW (\widehat{\lim}\star\widehat{\win_\Gamma}) \times \widehat{\Det_\Gamma}$, and the claim follows since $\lim \equivW \widehat{\lim}$.
\end{proof}
\end{theorem}

A further improvement on the techniques in \cite{leroux3} were provided by the authors in \cite{paulyleroux2,paulyleroux2-arxiv}. These techniques in particular suffice to prove the existence of subgame-perfect equilibria in antagonistic games (this implies two players and finitely many outcomes).

\begin{theorem}
\label{theo:thm19}
$\spe_\Gamma\leqW \widehat{\win_\Gamma} \times \widehat{\Det_\Gamma}$
\begin{proof}
Like in the proof of Theorem~\ref{theo:construct-NE} let us ask at every
node who wins the derived games and ask for a corresponding Nash
equilibrium. These two pieces of information together provide the existential
witness needed in the third condition of~\cite[Lemma 16]{paulyleroux2-arxiv}, so by the recursive construction in the proof of this lemma, its
conclusion follows computably. Let us invoke this conclusion once for each
of the two players and combine the obtained strategies into a strategy
profile, which is a subgame-perfect equilibrium by determinacy
assumptions. (See~\cite[Lemma 17]{paulyleroux2-arxiv} for the details.)
\end{proof}
\end{theorem}

\section{Deciding the winner and finding Nash equilibria}
The results in Section \ref{sec:det} show that for many concrete examples of $\Gamma$, the problem $\Det_\Gamma$ is inherently multivalued, i.e.~not equivalent to any functions between admissible spaces. On the other hand, the upper bounds provided in Section \ref{sec:transfer} all include $\win_\Gamma$, which is of course single-valued. In the current section, we will explore some converse reductions, from deciding the winner to finding Nash equilibria. This generally requires some (rather tame) requirements on the pointclasses involved.
\begin{lemma}
\label{lemma:wintone}
Let $\Gamma$ be obtained by $\Gamma_1$ by first closing under finite union, rescaling and union with clopens; and then adding complements. Then:
\[\win_{\Gamma_1}^* \leqW \NE_\Gamma^{ap}\]
\begin{proof}
We first informally describe the construction. Given $n$ win/lose games, the first player starts by announcing which of these games she believes she can win. Then the second player can choose one of the listed games to play. If the first player did not claim any winnable games, the game ends and the outcome is $0$. If the first player claimed to be able to win $k$ out of $n$ games, then the outcomes of the games subsequently chosen by the second player are scaled up to $k,-k$. Thus, the first player has every reason to list precisely those games she can actually win: If she would not list a game she could win, she trades payoff $k-1$ for payoff $k$. If she lists a game she cannot win, the second player will subsequently chose and win it, and then the first player is punished by $-k$.

The following depicts the construction in case of two input games:

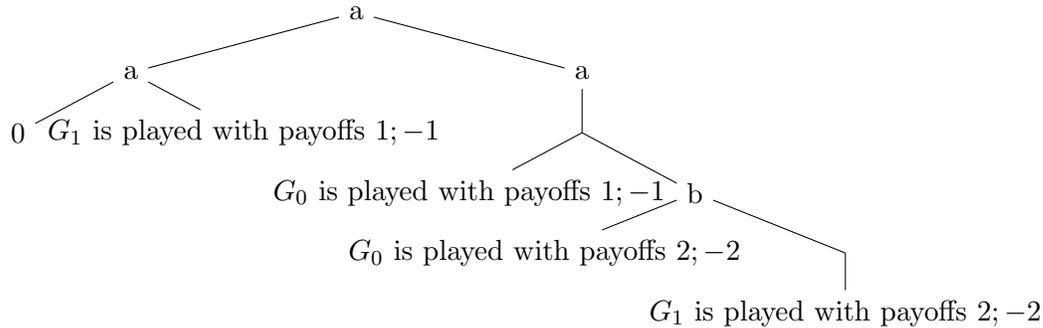
\begin{tikzpicture}[level distance=8mm]
\node{a}[sibling distance=60mm]
	child{node{a}[sibling distance=30mm]
		child{node{0}}
		child{node{$G_1$ is played with payoffs $1;-1$}}
	}
	child{node{a}[sibling distance=30mm]
		child{child{node{$G_0$ is played with payoffs $1;-1$}}
			child{node{b}[sibling distance=40mm]
		child{node{$G_0$ is played with payoffs $2;-2$}}
		child{child{node{$G_1$ is played with payoffs $2;-2$}}}}
	}};
\end{tikzpicture}

It remains to argue that the resulting game actually is a valid input to $\NE_\Gamma^{ap}$. We need to show that any upper set of outcomes is associated with a set of plays belonging to $\Gamma$. For $k > 0$, the upper set is a finite union of rescaled sets from $\Gamma_1$, which by assumption is a member of $\Gamma$. For $k = 0$, to the former we add the clopen set of plays resulting in $0$. For $k < 0$, we add additional clopens for those subgames played with stakes less than $k$.
\end{proof}
\end{lemma}

\begin{lemma}
\label{lemma:adjoinws}
Let $\Gamma$ be closed under taking unions with $\Gamma_1$ and $\overline{\Gamma_1}$. Then:
\[\NE_\Gamma^{ap} \times \textrm{FindWS}_{\Gamma_1} \equivW \NE_\Gamma^{ap}\]
\begin{proof}
As clearly $1 \leqW \textrm{FindWS}_\Gamma$, we only need to show $\NE_\Gamma^{ap} \times \textrm{FindWS}_\Gamma \leqW \NE_\Gamma^{ap}$. Let $G_0$ be the input game to $\NE_\Gamma^{ap}$ on the left, and $G_1$ be the input game to $\textrm{FindWS}_\Gamma$. If $\Omega$ is the (twice ordered) set of outcomes used in $G_0$, then we use the outcome set $\Omega \cup \{0,1\}$ for the newly constructed game, and extend the preferences by $0 \prec_a o \prec_a 1$ and $1 \prec_b o \prec_b 0$ for each $o \in \Omega$. The game tree looks as follows: The second player can choose whether to play in $G_1$ for outcomes $0$ (second player wins) and $1$ (second player loses), or to play in $G_0$ for the original outcomes:

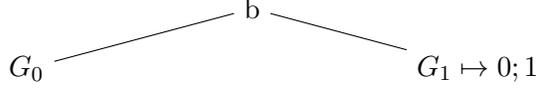
\begin{tikzpicture}[level distance=8mm]
\node{b}[sibling distance=60mm]
	child{node{$G_0$}[sibling distance=30mm]}
	child{node{$G_1 \mapsto 0;1$}[sibling distance=30mm]};
\end{tikzpicture}

In any Nash equilibrium of this game the second player is choosing to play in $G_0$, both players play a Nash equilibrium inside $G_0$, and the first player is using a winning strategy inside $G_1$. Thus, all desired information can be recovered.

In order for the constructed game to be valid for $\NE_\Gamma^{ap}$, we need that any upper set from $\Gamma$ is closed under union with a winning set of player 1.
\end{proof}
\end{lemma}

\begin{corollary}
\label{corr:nicegammalowerbound}
Let $\Gamma_1$ and $\Gamma$ simultaneously satisfy the criteria of the two preceding lemmata. Then:
\[\textrm{FindWS}_{\Gamma_1}^* \times \textrm{Win}_{\Gamma_1}^* \leqW \NE_{\Gamma}^{ap}\]
\end{corollary}

Unfortunately, the restrictions on $\Gamma$, $\Gamma_1$ in place in Lemma \ref{lemma:adjoinws} (and subsequently Corollary \ref{corr:nicegammalowerbound} are too strong for the application we have in mind. The result of Corollary \ref{corr:nicegammalowerbound} can be obtained with weaker conditions though:

\begin{lemma}
\label{lemma:nicegammalowerbound}
Let $\Gamma$ be obtained by $\Gamma_1$ by first closing under finite union, rescaling and union with clopens; and then adding complements. Then:
\[\textrm{FindWS}_{\Gamma_1}^* \times \win_{\Gamma_1}^* \leqW \NE_\Gamma^{ap}\]
\begin{proof}
The reduction directly combines the constructions in Lemma \ref{lemma:wintone} and Lemma \ref{lemma:adjoinws}. We only need to argue that the weaker condition on $\Gamma_1$ and $\Gamma$ suffices to have the valuation $\Gamma$-measurable in the resulting game. For this, note that the same reasoning as in Lemma \ref{lemma:wintone} applies, with the addition of a $\Gamma_1$ set above the other outcomes, and a $\overline{\Gamma}_1$ set below.
\end{proof}
\end{lemma}

If we have access to subgame perfect equilibria (and are in a context where they are guaranteed to exist), then we can even decide the winner of countably many games in parallel:

\begin{lemma}
\label{lemma:winspe}
Let $\Gamma_1$ contain the closed sets and be closed under finite unions and the operation $(A_n)_{n \in \mathbb{N}} \mapsto \left (\{0^\mathbb{N}\} \cup \bigcup_{n \in \mathbb{N}} 0^n1A_n \right )$. Let $\Gamma$ be obtained from $\Gamma_1$ by closing under complements. Then:
\[\widehat{\win_{\Gamma_1}} \leqW \spe_{\Gamma}\]
\begin{proof}
The input to $\widehat{\win_{\Gamma_1}}$ is a sequence of games $(G_i)_{i \in \mathbb{N}}$ with payoffs $0$ and $1$. From these, we shall construct a single game $G$ with payoffs $0$, $1$ and $\frac{1}{2}$. In $G$, the first player can move right as often as he desires. If he moves left once after having gone right $n$ times, then he is faced with the choice of moving left again and then playing $G_n$ against the second player, or to move right to receive a payoff of $\frac{1}{2}$. If player $1$ always moves right, he receives a payoff of $1$, as in the picture below.

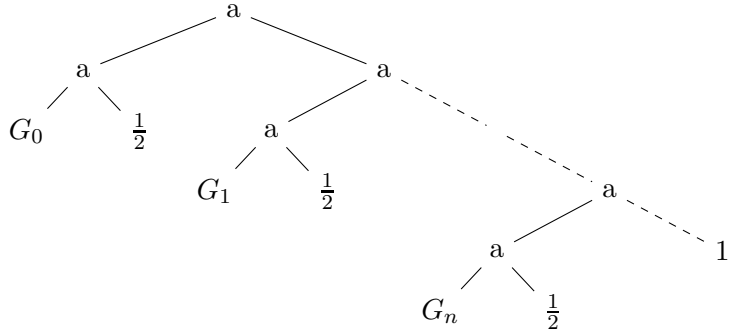
\begin{tikzpicture}[level distance=8mm]
\node{a}[sibling distance=40mm]
	child{node{a}[sibling distance=15mm]
		child{node{$G_0$}}
		child{node{$\frac{1}{2}$}}
	}
	child{node{a}[sibling distance=30mm]
		child{node{a}[sibling distance=15mm]
			child{node{$G_1$}}
			child{node{$\frac{1}{2}$}}
		}
		child{node{}[sibling distance=30mm] edge from parent [dashed]
			child{node{} edge from parent [draw = none]}
			child{node{a}[sibling distance=30mm]
				child{node{a}[sibling distance=15mm]edge from parent [solid]
					child{node{$G_n$}}
					child{node{$\frac{1}{2}$}}
				}
				child{node{$1$}}
			}
		}
	}
;
\end{tikzpicture}

In a subgame perfect equilibrium, the first player has to chose optimally when deciding between playing $G_n$ and receiving a guaranteed payoff of $\frac{1}{2}$ -- which means playing $G_n$ iff he can win it. Thus, any subgame perfect equilibrium of $G$ allows us to decide the winner of each $G_n$.

It only remains to show that the outcome sets of $G$ fall into $\Gamma$. For each $i \in \mathbb{N}$, let $U_i \in \Gamma_1$ be the winning set of the first player in $G_i$. Then the preimage of $\{1\}$ is $\{1\}^\mathbb{N} \cup \bigcup_{i \in \mathbb{N}} 1^i00U_i$, thus in $\Gamma_1$. The preimage of $\{\frac{1}{2}\}$ is an open set, and thus in $\Gamma$ together with the preimage of $\{1, \frac{1}{2}\}$. As $\Gamma$ is closed under complements, it then contains all relevant preimages.
\end{proof}
\end{lemma}

\section{General games with concrete pointclasses}
\label{sec:instanciate}
The general constructions put together with the classifications for specific pointclasses allow us to obtain some concrete Weihrauch degrees. First, we shall see that moving from a win/lose game with closed and open outcomes to a two-player game with several outcomes just complicates the operation of finding Nash equilibria by finitely many uses of $\lpo$ in parallel:

\begin{theorem}
\label{theo:concreteopenclosed}
$\NE^{ap}_{\mathcal{O}\cup\mathcal{A}} \equivW \C_\Cantor \times \lpo^*$
\begin{proof}
For the reduction $\NE^{ap}_{\mathcal{O}\cup\mathcal{A}} \leqW \C_\Cantor \times \lpo^*$, instantiate Theorem \ref{theo:equitransfer} with the results from Theorem \ref{theo:openclosedne} and Proposition \ref{prop:openclosedwin}.

For the other direction, note that $\textrm{FindWS}_{\mathcal{A}} \equivW \textrm{FindWS}_{\mathcal{O}\cup \mathcal{A}} \equivW \C_\Cantor$ as in Theorem \ref{theo:openclosedne}; and that $\Gamma_1 := \mathcal{A}$ and $\Gamma := \mathcal{O} \cup \mathcal{A}$ satisfy the requirements of Lemma \ref{lemma:nicegammalowerbound}, which then provides the desired result.
\end{proof}
\end{theorem}

The result can actually be strengthened into the following (by noting that the second game constructed in Lemma \ref{lemma:diffhierconstruction} is always won by the first player):

\begin{theorem}
$\NE^{ap}_{\mathfrak{D}_{n+1} \cup \overline{\mathfrak{D}_{n+1}}} \equiv_W \C_{\Cantor}^{[n]} \times \left (\lpo^{[n]} \right )^*$
\end{theorem}

If one wishes to have subgame-perfect equilibria instead of mere Nash equilibria, then countably many uses of $\lpo$ become necessary, and the problem becomes equivalent to $\lim$. Note that as long as there are at least three distinct outcomes, the number of outcomes has no further impact on the Weihrauch degree (due to the nature of the construction used to prove Lemma \ref{lemma:winspe})-- unlike the situation in Theorem \ref{theo:concreteopenclosed}, where the number of outcomes is related to the number of times that $\lpo$ is used.

\begin{theorem}
$\spe_{\mathfrak{D}_{n} \cup \overline{\mathfrak{D}_{n}}} \equivW \lim^{(n)}$
\begin{proof}
For $\spe_{\mathfrak{D}_n \cup \overline{\mathfrak{D}_n}} \leqW \lim^{(n)}$, instantiate Theorem \ref{theo:thm19} with the results from Theorem \ref{theo:differencehierarchy}, and note that $\widehat{\lpo^{(n)}} \equivW \lim^{(n)}$ and $\C_\Cantor^{[n]} \leqW \lim^{(n+1)}$.

For the other direction, we use Lemma \ref{lemma:winspe} (applicable by Corollary \ref{corr:diffhierpluspoint}) together with Proposition \ref{prop:openclosedwin}.
\end{proof}
\end{theorem}

Regarding Theorem \ref{theo:construct-NE}, we do not (yet?) have matching lower bounds for any particular pointclass. The gap is exemplified by the following:

\begin{corollary}
$\C_\Cantor \times \lpo^* \leqW \NE_{\mathcal{O}\cup\mathcal{A}} \leqW \lim$
\begin{proof}
The first reduction follows by Theorem \ref{theo:concreteopenclosed}. The second reduction follows by instantiating Theorem \ref{theo:construct-NE} with the results from Theorem \ref{theo:openclosedne} and Proposition \ref{prop:openclosedwin}.
\end{proof}
\end{corollary}

\section{Conclusions and Outlook}
With Theorem \ref{theo:differencehierarchy}, we have shown that the computational strength of determinacy provides a tight connection between the difference hierarchy and the Borel hierarchy (in form of Corollary \ref{corr:differencehierarchy3}). Note that winning sets from the difference hierarchy correspond to Boolean combinations of reachability and safety conditions. Corollary \ref{corr:differencehierarchy2} then provides an upper bound and a worst case for corresponding decidability questions for logic. Theorem \ref{theo:differencehierarchy} also shows that the computational powers of the players required to find a winning strategy vastly exceeds the computational power required to determine the outcome, thus casting doubt on the adequateness of winning strategies (or Nash equilibria) as adequate solution concepts for infinite sequential games\footnote{Then of course, infinite sequential games could justifiably be deemed unrealistic anyway.}.

The results in Section \ref{sec:transfer} contrasted with those in Section \ref{sec:instanciate} essentially show that the proofs in \cite{leroux3,leroux4,paulyleroux2,paulyleroux2-arxiv} are not too wasteful from a constructive perspective -- i.e.~the constructions employed are not far less constructive than the theorems proven with them.

There are several immediate avenues for extending the work presented here: The restriction to finite action sets (i.e.~finitely branching trees) can mostly be lifted without a significant impact on the proof techniques. Note though that the concrete Weihrauch degrees would change drastically, as in Theorem \ref{theo:openclosedne} we would need to replace $\C_\Cantor$ by $\C_\Baire$, with the latter residing in a less explored part of the Weihrauch lattice. The results in \cite{paulyleroux2,paulyleroux2-arxiv} are more general than covered here, too (with the same proof complexity). A notion of a product of sequential games could, similarly to the use of a products of bimatrix games in \cite{paulyincomputabilitynashequilibria} (see also \cite{pauly-xiang,pauly-xiang-arxiv}), be used to obtain some results on the products on the corresponding Weihrauch degrees. As in \cite{paulyleroux2,paulyleroux2-arxiv}, one could extend Theorem \ref{theo:differencehierarchy} to games with real-valued payoff functions of prescribed \emph{level} (introduced by \name{Hertling} in \cite{hertling,hertling4}, see also \cite{paulyreducibilitylattice,debrecht5}).

The study of the strength of determinacy for particular pointclasses in reverse mathematics presumably offers further proofs adaptable into the framework of Weihrauch reducibility, e.g.~\cite{nemoto2,montalban,eguchi}.

Another avenue to explore is the connection to the \emph{basic games} introduced by \name{Lachlan} \cite{lachlan} to formalize some meta-observations on the study of recursively enumerable sets. \name{Kummer} showed that the basic games are equivalent to the computable Gale Stewart games in the difference hierarchy over $\Sigma^0_2$ \cite{kummer2}. A natural restriction of the basic games can easily be shown to produce computable Gale-Stewart games in the difference hierarchy over $\Sigma^0_1$ as studied in this paper. This raises the question whether here another equivalence can be obtained, which would then allow us the transfer of our classifications to the setting of basic games\footnote{This question was originally raised by an anonymous referee.}.

A precise understanding of the Weihrauch degree of determinacy for various pointclasses is related to determinacy questions of games with transfinite length (as studied in e.g.~\cite{loewe,fraker}). For example, Corollary \ref{corr:differencehierarchy3} together with Borel determinacy implies that games of length $\omega + \omega$ and winning sets from the finite levels of the difference hierarchy are determined.

Further afield, understanding the Weihrauch degrees of determinacy principles may be a contribution to the development of descriptive set theory in computational/category-theoretical terms as suggested in \cite{pauly-descriptive}. In order to deal with determinacy beyond the Borel sets, one may have to adopt the \emph{extended Weihrauch-degrees} recently suggested by \name{Bauer} and \name{Yoshimura} \cite{yoshimura2} as framework.

\bibliographystyle{eptcs}
\bibliography{../../spieltheorie}

\section*{Acknowledgements}
This work benefited from the Royal Society International Exchange Grant IE111233 and the Marie Curie International Research Staff Exchange Scheme \emph{Computable
Analysis}, PIRSES-GA-2011- 294962.

\end{document}